\documentclass[superscriptaddress,twocolumn,pre]{revtex4-1}

\usepackage{amsmath}
\usepackage{graphicx,color}

\newcommand{\be}{\begin{equation}}
\newcommand{\ee}{\end{equation}}
\newcommand{\bea}{\begin{eqnarray}}
\newcommand{\eea}{\end{eqnarray}}

\begin{document}
\sloppy

%-title page-%

\title{A simple model of universe with a polytropic equation of state}

\author{Pierre-Henri Chavanis}
\email{chavanis@irsamc.ups-tlse.fr}
\affiliation{Laboratoire de Physique Th\'eorique (IRSAMC), CNRS and UPS, Universit\'e de Toulouse, France}

\begin{abstract}

We construct a simple model of universe with a generalized equation of state $p=(\alpha +k\rho^{1/n})\rho c^2$ having a linear component $p=\alpha\rho c^2$ and a polytropic component $p=k\rho^{1+1/n}c^2$. For $\alpha=1/3$, $n=1$ and $k=-4/(3\rho_P)$, where $\rho_P=5.16\, 10^{99}\, {\rm g}/{\rm m}^3$ is the Planck density, this equation of state provides a model of the early universe without singularity describing the transition between the pre-radiation era and the radiation era. The universe starts from $t=-\infty$ but, when $t<0$, its size is less than the Planck length $l_P=1.62\, 10^{-35}\, {\rm m}$. The universe undergoes an inflationary expansion that brings it to a size $a_1=2.61\, 10^{-6}\, {\rm m}$ on a timescale of a few Planck times $t_P=5.39\, 10^{-44}\, {\rm s}$. When $t\gg t_P$, the universe decelerates and enters in the radiation era. For $\alpha=0$, $n=-1$ and $k=-\rho_{\Lambda}$, where $\rho_{\Lambda}=7.02\, 10^{-24}\, {\rm g}/{\rm m}^3$ is the cosmological density, this equation of state describes the transition from a  decelerating universe dominated by baryonic and dark matter to an accelerating universe  dominated by dark energy (second inflation). The transition takes place at a size $a_2=8.95\, 10^{25}{\rm m}$ corresponding to a time of the order of the cosmological time $t_{\Lambda}=1.46\, 10^{18}\, {\rm s}$. The present universe turns out to be just at the transition ($t_0\sim t_{\Lambda})$. This polytropic model reveals a nice ``symmetry'' between the early  and late evolution of the universe, the cosmological constant $\Lambda$ in the late universe playing a role similar to the Planck constant $\hbar$ in the early universe. We interpret the cosmological constant as a fundamental constant of nature describing the ``cosmophysics'' just like the Planck constant describes the microphysics. The Planck density and the cosmological density represent fundamental upper and lower bounds differing by ${122}$ orders of magnitude. The cosmological constant ``problem'' may be a false problem.

\end{abstract}

\maketitle

According to contemporary cosmology, the present energy content of the universe is composed of approximately $5\%$ baryonic matter, $20\%$ dark matter and $75\%$ dark energy \cite{bt}. The expansion of the universe began in a tremendous inflationary burst driven by vacuum energy. Between $10^{-35}$ and $10^{-33}$ seconds after the Big Bang, the universe expanded by a factor $10^{30}$ \cite{linde}. The universe then entered in the radiation era and, when the temperature cooled down below approximately $10^3\, {\rm K}$, in the matter era \cite{weinberg}. At present, it undergoes an accelerated expansion  presumably due to the cosmological constant or to some form of dark energy \cite{cst}. This corresponds to a second period of inflation. Despite the success of the standard model, the nature of dark matter, dark energy, and the evolution of the very early universe (pre-radiation era), remains very mysterious and leads to many speculations. In this Letter, we propose to parameterize this indetermination by the concept of an equation of state. We show that the early and the late evolution of the universe can be described ``symmetrically'' by two polytropic equations of state with index $n=1$ and $n=-1$, respectively. Polytropic equations of state play an important role in astrophysics \cite{chandra,st}, statistical physics \cite{tsallisbook}, and mathematical biology \cite{murray}, and they may also be useful in cosmology. Our results are consistent with the standard $\Lambda$CDM model but refine it by removing the primordial singularity.

We assume that the universe is isotropic and homogeneous at large scales and contains a uniform perfect fluid of energy density $\epsilon(t)=\rho(t) c^2$ and pressure $p(t)$. We also assume that the universe is flat in agreement with observations of the cosmic microwave background (CMB) \cite{bt}. It that case, the Einstein equations reduce to
\begin{equation}
\label{a1}
\frac{d\rho}{dt}+3\frac{\dot a}{a}\left (\rho+\frac{p}{c^2}\right )=0,
\end{equation}
\begin{equation}
\label{a2}
H^2=\left (\frac{\dot a}{a}\right )^2=\frac{8\pi G}{3}\rho,
\end{equation}
where $a(t)$ is the scale factor (``radius'' of the universe) and $H=\dot a/a$ is the Hubble parameter. These are the well-known Friedmann equations describing a non-static universe \cite{weinberg}. The first equation can be viewed as an ``equation of continuity''. For a given barotropic equation of state $p=p(\rho)$, it determines the relation between the density and the scale factor. Then, the temporal evolution of the scale factor is given by Eq. (\ref{a2}). We have not written the cosmological constant $\Lambda$ because its effect will be taken into account in the equation of state. We will also need the thermodynamical equation
\begin{equation}
\label{a3}
\frac{dp}{dT}=\frac{1}{T}(\rho c^2+p),
\end{equation}
which can be derived from the first principle of thermodynamics \cite{weinberg}. For a given barotropic equation of state $p=p(\rho)$, this equation can be integrated to obtain the relation $T=T(\rho)$ between the temperature and the density. To close the system of equations, we have to specify an equation of state. We consider a generalized equation of state of the form
\begin{equation}
\label{a4}
p=(\alpha \rho+k\rho^{1+1/n}) c^2.
\end{equation}
This is the sum of a standard linear equation of state $p=\alpha\rho c^2$ describing a universe dominated by radiation ($\alpha=1/3$) or pressureless matter ($\alpha=0$), and a polytropic equation of state $p=k\rho^{\gamma} c^2$, where $k$ is the polytropic constant and $\gamma=1+1/n$ the polytropic index.  An exhaustive study of this equation of state, considering all possible cases, is given in \cite{prep}. In this Letter, we describe two specific models of particular physical interest.

We first consider the case $\alpha=1/3$, $n=1$ and $k=-4/(3\rho_P)$, which provides a model of the early universe without singularity. The equation of state (\ref{a4}) can be rewritten as
\begin{equation}
\label{a5}
p=\frac{1}{3}(1-4\rho/\rho_P)\rho c^2.
\end{equation}
The linear part describes the radiation and the quadratic part may be due to Bose-Einstein condensates (BECs) with attractive self-interaction \cite{c4}, or have another origin. The continuity equation (\ref{a1}) can be integrated into
\begin{equation}
\label{a6}
\rho=\frac{\rho_P}{(a/a_1)^{4}+ 1},
\end{equation}
where $a_1$ is a constant of integration.

{\it Pre-radiation era:} When $a\ll a_1$, the density is approximately constant: $\rho\simeq \rho_P$. Since this solution is expected to describe the very early universe, it is natural to identify the constant $\rho_P$ with the Planck density which may represent a fundamental upper bound for the density. A constant value of the density gives rise to a phase of early inflation. From the Friedmann equation (\ref{a2}), we find that the scale factor increases exponentially rapidly with time as
\begin{equation}
\label{a7}
a(t)\sim l_P e^{({8\pi}/{3})^{1/2}t/t_P}.
\end{equation}
We have defined the time $t=0$ such that $a(0)$ is equal to the Planck length $l_P$. This universe exists at any time in the past ($a\rightarrow 0$ and $\rho\rightarrow \rho_P$ for $t\rightarrow -\infty$), so there is no primordial singularity.

{\it Radiation era:} When $a\gg a_1$, we get $\rho/\rho_P\sim (a_1/a)^4$. When the density is ``low'', the equation of state (\ref{a5}) reduces to $p=\rho c^2/3$ which describes the radiation era. The conservation of $\rho_{rad}\,  a^4$ implies that $\rho_P a_1^4=\rho_{rad,0}\, a_0^4$ where $\rho_{rad,0}$ is the present density of radiation and $a_0=c/H_0=1.32\, 10^{26}\, {\rm m}$ the present distance of cosmological horizon determined by the Hubble constant $H_0=2.27\, 10^{-18}\, {\rm s}^{-1}$. Writing $\rho_{rad,0}=\Omega_{rad,0}\rho_0$  where $\rho_0=9.20\, 10^{-24}\, {\rm g}/{\rm m}^3$ is the present density of the universe given by Eq. (\ref{a2}), and $\Omega_{rad,0}=8.48\, 10^{-5}$, we obtain
\begin{equation}
\label{a8}
a_1=2.61\, 10^{-6}\, {\rm m}.
\end{equation}
This scale is  intermediate between the Planck length $l_P$ and the present size of the universe $a_0$ ($a_1/l_P=1.61\, 10^{29}$ and $a_1/a_0=1.97\, 10^{-32}$). It gives the typical size of the universe at the end of the inflationary phase (or  at the beginning of the radiation era). When $a\gg a_1$, the Friedmann equation (\ref{a2}) yields $a/a_1\sim ({32\pi}/{3})^{1/4}({t}/{t_P})^{1/2}$ and $\rho/\rho_P\sim ({3}/{32\pi})({t_P}/{t})^2$.

{\it The general solution:} The general solution of the Friedmann equation (\ref{a2}) with Eq. (\ref{a6}) is
\begin{eqnarray}
\label{a9}
\sqrt{(a/a_1)^{4}+1}-\ln \left (\frac{1+\sqrt{(a/a_1)^{4}+1}}{(a/a_1)^{2}}\right )\nonumber\\
=2\left (\frac{8\pi}{3}\right )^{1/2} \frac{t}{t_P}+C,
\end{eqnarray}
where $C\simeq -134$ is a constant of integration determined such that $a=l_P$ at $t=0$. The universe is accelerating when $a<a_c$ (i.e.
$\rho>\rho_c$) and decelerating when $a>a_c$ (i.e.  $\rho<\rho_c$) where $\rho_c=\rho_P/2$ and $a_c=a_1$. The time $t_c$ at which the universe starts decelerating is given by ${t_c}=({3}/{32\pi})^{1/2}\lbrack \sqrt{2}-\ln(1+\sqrt{2})-C\rbrack t_P$.
This corresponds to the time at which the curve $a(t)$ presents an inflexion point. It turns out that this inflexion point coincides with $a_1$ so it also marks the end of the inflation ($t_c=t_1$).

{\it Evolution of the temperature:} The thermodynamical equation (\ref{a3}) with Eq. (\ref{a5}) can be integrated into
\begin{equation}
\label{a11}
T=T_P \left (\frac{15}{\pi^2}\right )^{1/4} \left (1- \frac{\rho}{\rho_P}\right )^{7/4}\left (\frac{\rho}{\rho_P}\right )^{1/4},
\end{equation}
\begin{equation}
\label{a12}
T=T_P \left (\frac{15}{\pi^2}\right )^{1/4} \frac{(a/a_1)^7}{\left\lbrack (a/a_1)^4+ 1\right\rbrack^2},
\end{equation}
where $T_P$ is a constant of integration. In the radiation era $\rho\ll\rho_P$, Eq. (\ref{a11}) reduces to  $\rho/\rho_P\sim (\pi^2/15)(T/T_P)^4$ which is the Stefan-Boltzmann law provided that $T_P$ is identified with the Planck temperature $T_P=1.42\, 10^{32}\, {\rm K}$. The temperature starts from $T=0$ at $t=-\infty$, increases exponentially rapidly during the inflation, reaches a maximum value $T_e$, and decreases algebraically as  $T/T_P\sim ({45}/{32\pi^3})^{1/4} ({t_P}/{t})^{1/2}$ during the radiation era. The point corresponding to the maximum temperature is
$\rho_{e}=\rho_P/8$, $a_{e}=7^{1/4}a_1$, $T_{e}=({7}/{8})^{7/4}({15}/{8\pi^2})^{1/4}T_P$. It is reached at a time $t_{e}=({3}/{32\pi})^{1/2}\lbrack \sqrt{8}-\ln ({1+\sqrt{8}})+\ln7/2-C\rbrack t_P$.

\begin{figure}[!h]
\begin{center}
\includegraphics[clip,scale=0.3]{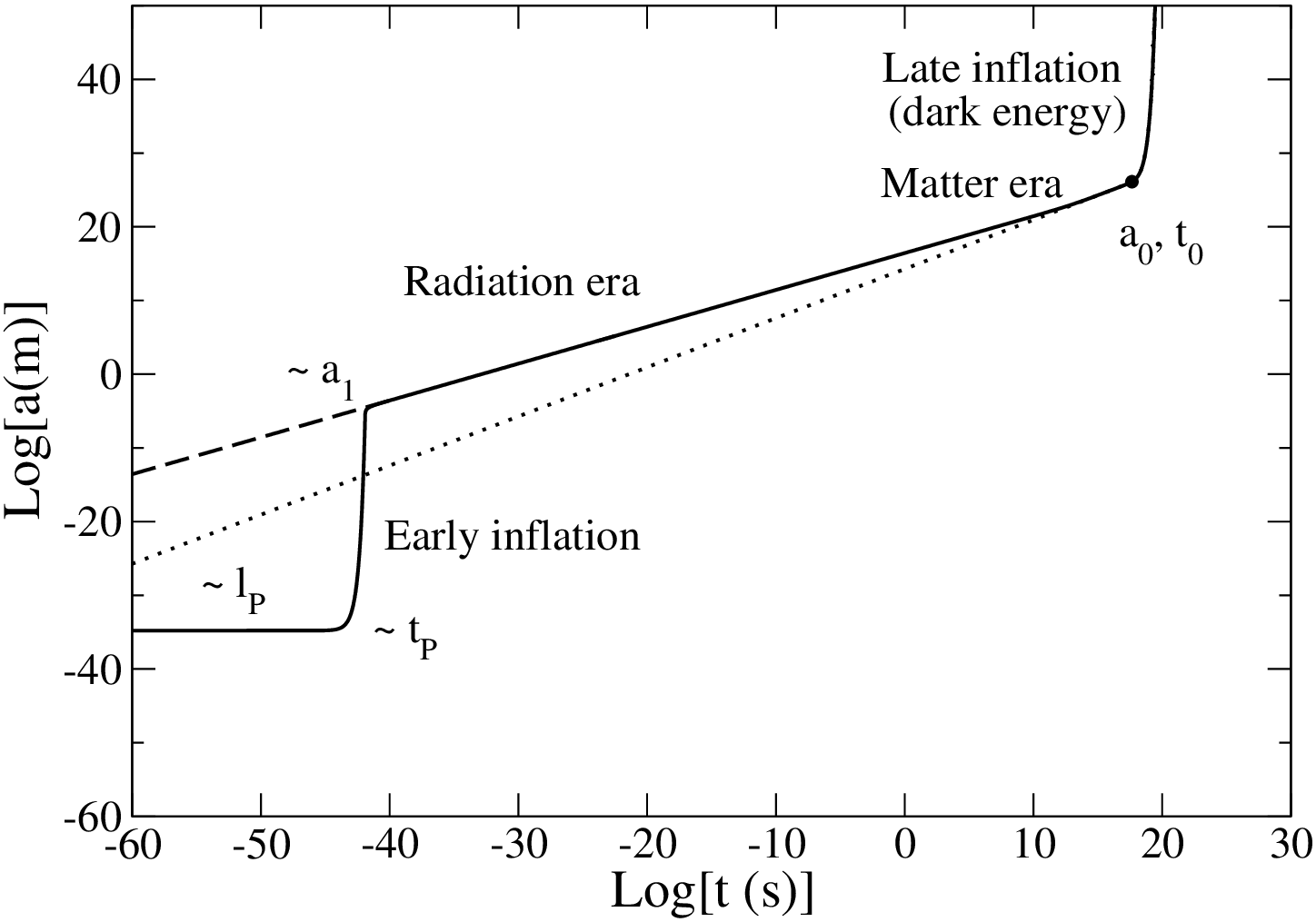}
\caption{Evolution of the scale factor $a$ as a function of time. There is an early inflation (due to the Planck density $\rho_P$) connecting the pre-radiation era to the radiation era. During the early inflation, the scale factor increases by $29$ orders of magnitude in less than $10^{-42}\, {\rm s}$. There is a late inflation (due to the cosmological density $\rho_{\Lambda}$) connecting the matter era to the dark energy era. A present, we live just at the transition (bullet).}
\label{taLOGLOGcomplet}
\end{center}
\end{figure}

\begin{figure}[!h]
\begin{center}
\includegraphics[clip,scale=0.3]{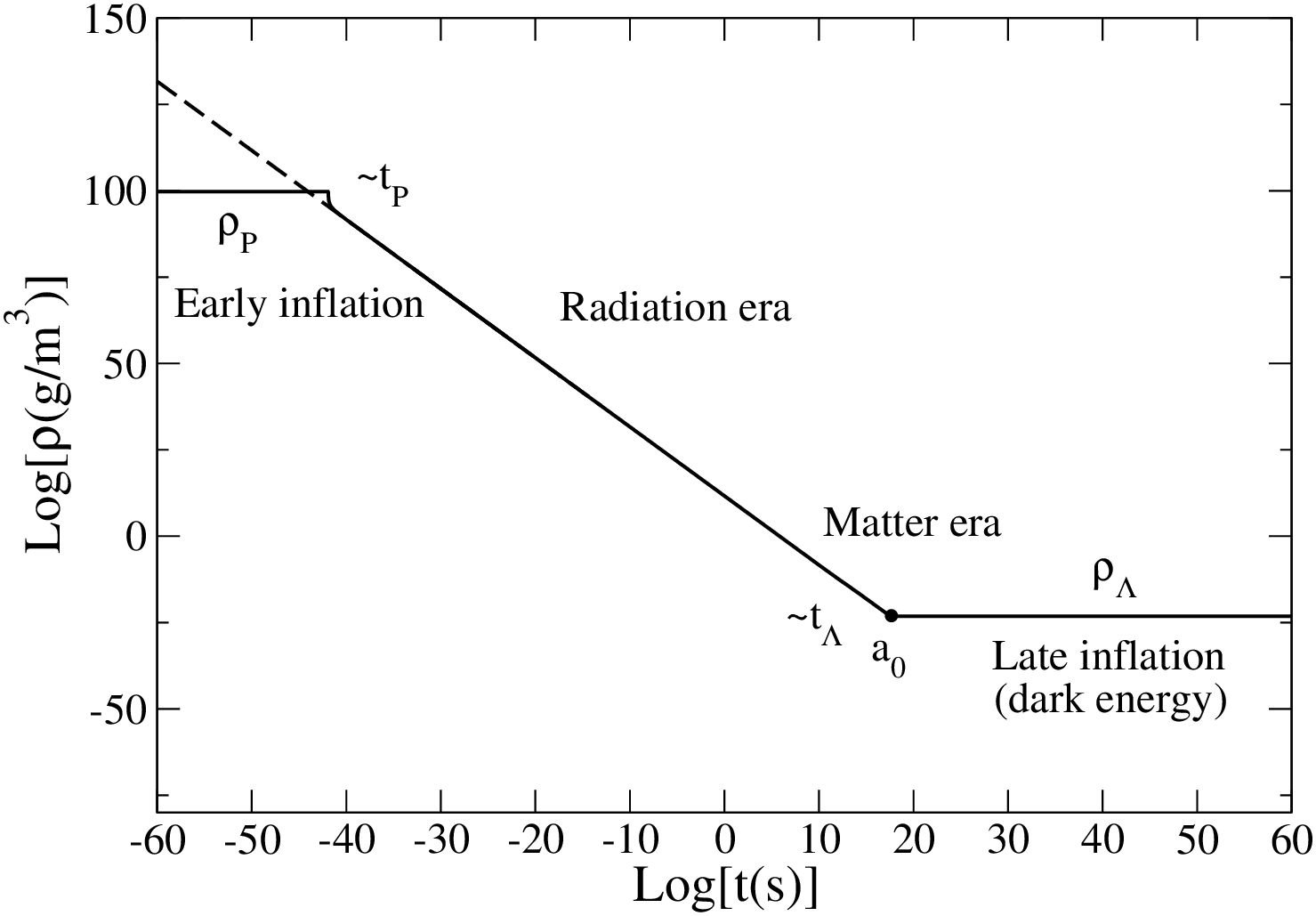}
\caption{Evolution of the density $\rho$ as a function of time. It varies between two bounds $\rho_P$ and $\rho_{\Lambda}$, fixed by fundamental constants of physics, producing two phases of inflation.}
\label{trhoLOGLOGcomplet}
\end{center}
\end{figure}

\begin{figure}[!h]
\begin{center}
\includegraphics[clip,scale=0.3]{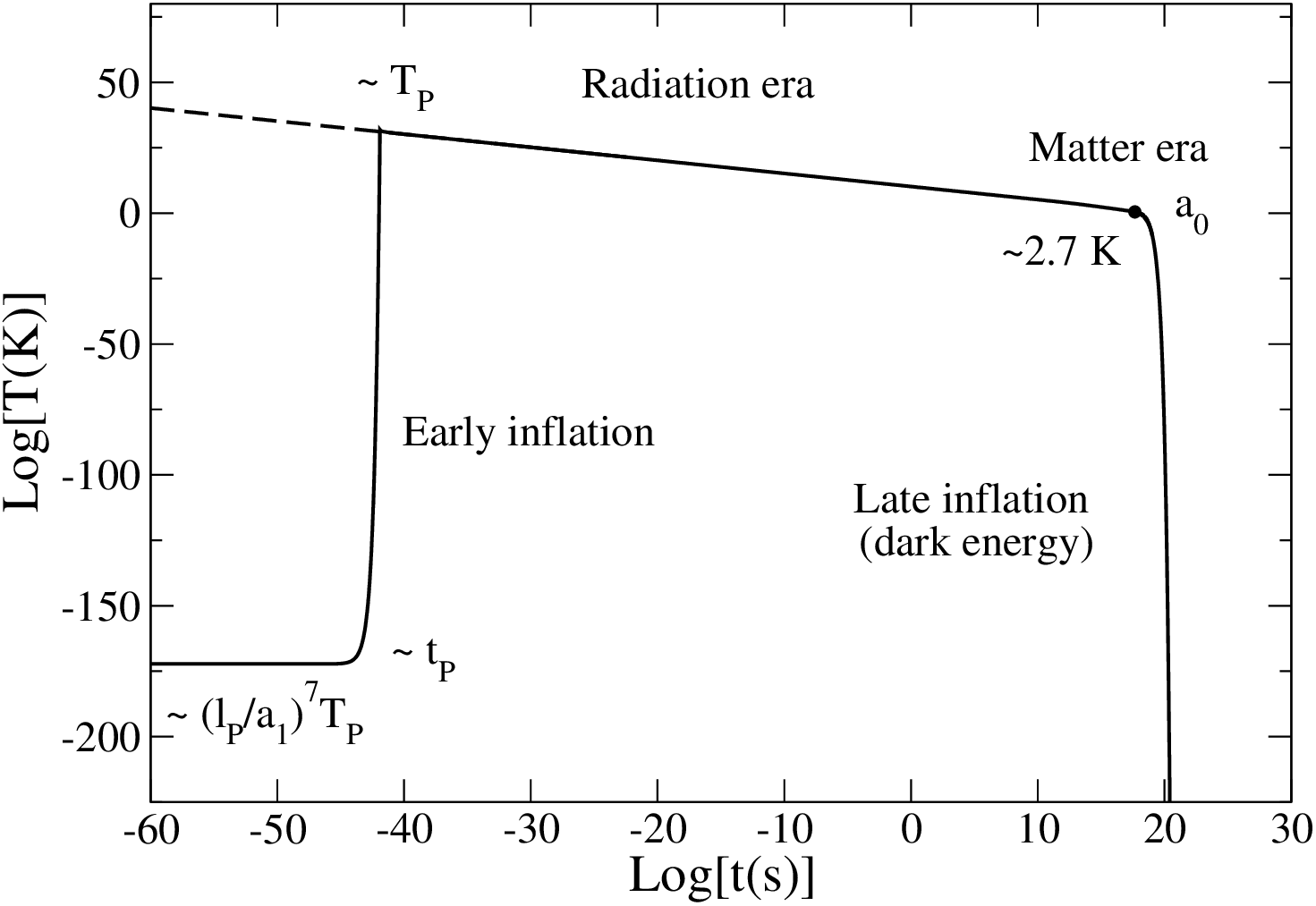}
\caption{Evolution of the temperature $T$ as a function of time. Before the early inflation, the universe is extremely cold ($T<10^{-173}\, {\rm K}$). During the early inflation, the temperature increases by $204$ orders of magnitude in less than $10^{-42}\, {\rm s}$. During the radiation era, the temperature decreases algebraically. At present, $T\simeq 2.7\, {\rm K}$ (bullet).}
\label{ttempLOGLOG}
\end{center}
\end{figure}

{\it Early universe:} The universe starts at  $t=-\infty$ with a vanishing radius $a=0$, a finite density $\rho=\rho_P=5.16\, 10^{99}\, {\rm g/m^3}$, and a vanishing temperature $T=0$. The universe exists at any time in the past and does not present any singularity. We define the ``original'' time $t=t_i=0$ as the time at which the radius of the universe is equal to the Planck length. Thus $a_i=l_P=1.62\, 10^{-35}\, {\rm m}$. The corresponding density and temperature are $\rho_i\simeq \rho_P=5.16\, 10^{99}\, {\rm g}/{\rm m}^3$ and $T_i=3.91\, 10^{-205} \, T_P=5.54\, 10^{-173}\, {\rm K}$. We note that quantum mechanics regularizes the finite time singularity present in the standard Big Bang theory. This is similar to finite size effects in second order phase transitions (the Big Bang theory is recovered for $\hbar=0$) \cite{prep}. We also note that the universe is very cold at $t=t_i=0$, unlike what is predicted by the Big Bang theory (a naive extrapolation of the law $T\propto t^{-1/2}$ leads to $T(0)=+\infty$). The universe first undergoes a phase of inflation during which its radius and temperature increase exponentially rapidly while its density remains approximately constant. The inflation ``starts'' at $t_i=0$ and ends at
$t_1=23.3\, t_P=1.25\, 10^{-42}\, {\rm s}$. During this very short lapse of time,  the radius of the universe grows from $a_i=l_P=1.62\, 10^{-35}\, {\rm m}$ to $a_1=2.61\, 10^{-6}\, {\rm m}$, and the temperature grows from $T_i=3.91\, 10^{-205} \, T_P=5.54\, 10^{-173}\, {\rm K}$ to $T_1=0.278\, T_P=3.93\, 10^{31}\, {\rm K}$. By contrast, the density does not change significatively: It goes from  $\rho_i\simeq \rho_P=5.16\, 10^{99}\, {\rm g}/{\rm m}^3$ to $\rho_1=0.5\, \rho_P=2.58\, 10^{99}\, {\rm g/m^3}$. After the inflation, the universe enters in the radiation era and, from that point, we recover the standard model \cite{weinberg}. The radius increases algebraically as $a\propto t^{1/2}$  while the density and the temperature decrease algebraically as $\rho\propto t^{-2}$ and $T\propto t^{-1/2}$. The temperature achieves its maximum value $T_{e}=0.523\, T_P=7.40\, 10^{31}\, {\rm K}$ at $t=t_{e}=23.6\, t_P=1.27\, 10^{-42}\, {\rm s}$.  At that moment, the density is $\rho_{e}=0.125\, \rho_P=6.44\, 10^{98}\, {\rm g}/{\rm m}^3$ and the radius $a_{e}=1.63\, a_1=4.24\, 10^{-6}\, {\rm m}$. During the inflation, the universe is accelerating and during the radiation era it is decelerating. The transition takes place at a time $t_c=t_1=23.3\, t_P=1.25\, 10^{-42}\, {\rm s}$ coinciding with the end of the inflation ($a_c=a_1$). The evolution of the scale factor, density  and temperature as a function of time are represented in Figs. \ref{taLOGLOGcomplet}-\ref{ttempLOGLOG} in logarithmic scales.

{\it Scalar field}: The phase of inflation in the very early universe is usually described by a scalar field \cite{linde}. We have determined the ordinary scalar field (minimally coupled to gravity) leading to the equation of state (\ref{a5}) and found that it corresponds to a potential
\begin{equation}
\label{a14}
V(\psi)=\frac{1}{3}\rho_P c^2 \frac{\cosh^2\psi+2}{\cosh^4\psi}.
\end{equation}

We now consider the case $\alpha=0$, $n=-1$ and $k=-\rho_{\Lambda}$, which provides a model of the late universe dominated by dark matter and dark energy. The equation of state can be rewritten as
\begin{equation}
\label{a15}
p=-\rho_{\Lambda} c^2.
\end{equation}
The pressure has a constant negative value. It belongs to the family of the generalized Chaplygin model $p=-A/\rho^{a}$ with $a\ge -1$ (here $a=0$) \cite{chaplygin,cst}.  The continuity equation (\ref{a1}) can be integrated into
\begin{equation}
\label{a16}
\rho=\rho_{\Lambda}\left\lbrack \left (\frac{a_2}{a}\right )^{3}+ 1\right\rbrack,
\end{equation}
where $a_2$ is a constant of integration.

{\it Dark energy era:} When $a\gg a_2$, the density is approximately
constant: $\rho\simeq \rho_{\Lambda}$. Since this solution is expected
to describe the very late universe, it is natural to identify the constant
$\rho_{\Lambda}$ with the cosmological density
$\rho_{\Lambda}={\Lambda}/{8\pi G}=7.02\, 10^{-24}\, {\rm g}/{\rm
m}^3$ (dark energy) which may represent a
fundamental lower bound for the density. A constant value of the
density gives rise to a phase of late inflation. It is convenient to define
a cosmological time
$t_{\Lambda}=1/(G\rho_{\Lambda})^{1/2}=(8\pi/\Lambda)^{1/2}=1.46\,
10^{18}\, {\rm s}$ and a cosmological length $l_{\Lambda}=c
t_{\Lambda}=(8\pi c^2/\Lambda)^{1/2}=4.38\, 10^{26}\, {\rm m}$. These are the counterparts of the Planck scales for the late universe. From the Friedmann
equation (\ref{a2}), we find that the scale factor increases
exponentially rapidly with time as
\begin{equation}
\label{a17}
a(t)\sim l_{\Lambda} e^{(8\pi/3)^{1/2}(t-t_f)/t_{\Lambda}},
\end{equation}
with $t_{f}=0.708 t_{\Lambda}=1.03\, 10^{18}\, {\rm s}$ (see below). This exponential growth corresponds to the so-called de Sitter solution \cite{bt}.

{\it Matter era:} When $a\ll a_2$, we get $\rho/\rho_{\Lambda}\sim (a_2/a)^3$. Because of the smallness of $\rho_{\Lambda}$, the equation of state (\ref{a15}) can be approximated by $p\simeq 0$ which describes the pressureless matter era \cite{weinberg}. The conservation of $\rho_{m}\,  a^3$ implies that $\rho_{\Lambda} a_2^3=\rho_{m,0}\, a_0^3$. Using $\rho_{\Lambda}=\Omega_{\Lambda,0}\rho_0$  and $\rho_{m,0}=\Omega_{m,0}\rho_0$  with $\Omega_{\Lambda,0}=0.763$  (dark energy) and $\Omega_{m,0}=\Omega_{B,0}+\Omega_{DM,0}=0.237$ (baryons and dark matter), we obtain
$a_2/a_0=(\Omega_{m,0}/\Omega_{\Lambda,0})^{1/3}$ hence
\begin{equation}
\label{a18}
a_2=8.95\, 10^{25}\, {\rm m}.
\end{equation}
This can be rewritten as $a_2=0.204 l_{\Lambda}$. When $a\ll a_2$, the Friedmann equation (\ref{a2}) yields $a/a_2\sim (6\pi)^{1/3}(t/t_{\Lambda})^{2/3}$ and $\rho/\rho_{\Lambda}\sim (1/6\pi)(t_{\Lambda}/t)^2$ (Einstein-de Sitter solution).

{\it The general solution:} The general solution of the Friedmann equation (\ref{a2}) with Eq. (\ref{a16}) is
\begin{eqnarray}
\label{a19}
\frac{a}{a_2}=\sinh^{2/3}\left (\sqrt{6\pi}\frac{t}{t_{\Lambda}}\right ).
\end{eqnarray}
Its asymptotic behavior gives Eq. (\ref{a17}) with $t_f=({3}/{8\pi})^{1/2}\lbrack ({2}/{3})\ln2+\ln(l_{\Lambda}/a_2)\rbrack t_{\Lambda}$. The universe is decelerating when $a<a'_c$ (i.e. $\rho>\rho'_c$) and accelerating when $a>a'_c$ (i.e.  $\rho<\rho'_c$) where $a'_c=(1/2)^{1/3}a_2$ and $\rho'_c=3\rho_{\Lambda}$. The time $t'_c$ at which the universe starts accelerating is given by $t'_c=({1}/{6\pi})^{1/2}{\rm argsinh}(1/\sqrt{2})t_{\Lambda}$.
This corresponds to the time at which the curve $a(t)$ presents a second inflexion point.

{\it Late universe:} In the matter era, the radius increases  algebraically as $a\propto t^{2/3}$ while the density decreases algebraically as $\rho\propto t^{-2}$ (EdS). When $a\gg a_2$, the universe enters in the dark energy era. It undergoes a late inflation (de Sitter) during which its radius increases exponentially rapidly while  its density remains constant and equal to the cosmological density $\rho_{\Lambda}$. The transition takes place at $a_2=0.204 l_{\Lambda}=8.95\, 10^{25}\, {\rm m}$, $\rho_2=2\rho_{\Lambda}=1.40\, 10^{-23}{\rm g}/{\rm m}^3$, $t_2=0.203 t_{\Lambda}=2.97\, 10^{17}{\rm s}$. In the matter era, the universe is decelerating and in the dark energy era it is accelerating. The time at which the universe starts accelerating is $t_{c}=0.152 t_{\Lambda}=2.22\, 10^{17}\, {\rm s}$, corresponding to a radius $a_c=0.162 l_{\Lambda}=7.11\, 10^{25}{\rm m}$ and a density $\rho_c=3\rho_{\Lambda}=2.11\, 10^{-23}{\rm g}/{\rm m}^3$.

{\it Present universe:} The present size  of the universe $a_0=0.302 l_{\Lambda}=1.32\, 10^{26}{\rm m}$ is precisely of the order of the scale $a_2=0.204 l_{\Lambda}=8.95\, 10^{25}\, {\rm m}$ ($a_0=1.48 a_2$). Therefore, we live just at the transition between the matter era and the dark energy era (see bullets in Figs. \ref{taLOGLOGcomplet} and \ref{trhoLOGLOGcomplet}). The present density of the universe is $\rho_0=1.31\rho_{\Lambda}=9.20\, 10^{-24}{\rm g}/{\rm m}^3$ and the present value of the deceleration parameter is $q_0=(1-3\Omega_{\Lambda,0})/2=-0.645$.  The age of the universe is
$t_{0}=({1}/{6\pi})^{1/2}{\rm argsinh}\lbrack ({a_0}/{a_2})^{3/2}\rbrack t_{\Lambda}$. Numerically, $t_0=0.310t_{\Lambda}=4.54\, 10^{17}{\rm s}$.

{\it Scalar field:} In  alternative theories to the cosmological constant, the present-day acceleration of the universe (dark energy) is described by a scalar field called quintessence \cite{cst}.  We have determined the  potential of the quintessence field associated with the equation of state (\ref{a15}) and found that it is given by
\begin{equation}
\label{a21}
V(\psi)=\frac{1}{2}\rho_{\Lambda} c^2 (\cosh^2\psi+1), \qquad (\psi\ge 0).
\end{equation}
On the other hand, since $w=p/\rho c^2=-\rho_{\Lambda}/\rho$ is between $-1$ and $0$ ($w_0=-\Omega_{\Lambda,0}=-0.763$), the equation of state (\ref{a15}) can also be associated with a tachyon field \cite{cst}. We have found that it corresponds to a potential
\begin{equation}
\label{a22}
V(\psi)=\frac{\rho_{\Lambda} c^2}{\cos\psi}, \qquad (0\le \psi\le \pi/2).
\end{equation}

{\it Conclusion:} We have considered two simple polytropic equations of state with negative pressure and index $n=1$ and $n=-1$. These equations of state are relatively ``symmetric''. They describe respectively a phase of inflation in the early universe (due to the Planck density $\rho_P$) and in the late universe (due to the cosmological $\rho_{\Lambda}$). They also describe, in a unified manner,  the transition from the pre-radiation era to the radiation era in the early universe, and the transition from the matter era to the dark energy era in the late universe. In the early universe, this amounts to summing  the inverse Planck density and the inverse radiation density  ($1/\rho=1/\rho_P+1/\rho_{rad}$); in the late universe, this amounts to summing the  matter density and the cosmological density ($\rho=\rho_{m}+\rho_{\Lambda}$). The whole evolution of the universe can  be described by a single equation
\begin{equation}
\label{a23}
\frac{H}{H_0}=\sqrt{\frac{\Omega_{rad,0}}{(a/a_0)^4+(a_1/a_0)^4}+\Omega_{m,0}^*\left\lbrack (a_0/a)^3+(a_0/a_2)^3\right\rbrack},
\end{equation}
combining the merits of these two models (the asterisk means that the matter term must be introduced for $a\gg a_1$ in order to avoid a spurious divergence when $a\rightarrow 0$).  This equation can be written equivalently as
\begin{equation}
\label{a24}
\frac{H}{H_0}=\sqrt{\frac{\Omega_{rad,0}}{(a/a_0)^4+(a_1/a_0)^4}+\frac{\Omega_{m,0}^*}{(a/a_0)^3}
+\Omega_{\Lambda,0}},
\end{equation}
For $a_1=0$, we recover the standard $\Lambda$CDM model which presents a singularity at $t=0$ \cite{bt}. For $a_1\neq 0$, the initial singularity is  removed and the universe exists eternally in past and future (aioniotic universe). This model is the most natural non-singular symmetric solution of the cosmological Einstein equations \cite{prep}.

{\it Cosmological constant problem:} It is oftentimes argued that the dark energy density $\rho_{\Lambda}$ represents the vacuum energy density. Since the vacuum energy density is expected to be of the order of the Planck density $\rho_P\sim 10^{122} \rho_{\Lambda}$, this leads to the so-called cosmological constant problem \cite{weinbergcosmo}. Actually, as illustrated in Fig. \ref{trhoLOGLOGcomplet}, the Planck density and the cosmological density represent fundamental upper and lower density bounds acting in the early and late universe, respectively. Therefore, it is not surprising that they differ by ${122}$ orders of magnitude. We view the cosmological constant as a fundamental constant of nature describing the ``cosmophysics'' (late universe) in the same sense that the Planck constant describes the microphysics (early universe). Accordingly, the origin of the  dark energy density $\rho_{\Lambda}$ should not be sought in quantum mechanics, but in pure general relativity. In this sense, the cosmological constant ``problem'' may be a false problem. Of course, the origin of the cosmological constant still remains to be understood.

\end{document}